\documentclass[10pt,twocolumn]{IEEEtran}

\usepackage{amsmath,amssymb}
\usepackage{epsfig}
\usepackage{verbatim}
\usepackage{subfigure}
\usepackage[dvips]{color}
\usepackage{setspace}

\newcommand{\G}{{\mathcal G}}

\newtheorem{thm}{Theorem}[section]

\begin{document}
\title{\Large On Models for Multi-User Gaussian Channels with Fading}
\author{Rony El Haddad, Brian Smith and Sriram Vishwanath \footnote{This work supported by a grant from the Army Research Office and the National Science Foundation CAREER award.}
\thanks{The authors are with Wireless Networking and Communications Group,
Department of Electrical and Computer Engineering, The University of
Texas at Austin, Austin, TX 78712, USA (e-mail: bsmith,sriram@ece.utexas.edu;rony.elhaddad@mail.utexas.edu).}
}
\maketitle

\thispagestyle{empty}
\pagestyle{empty}
\begin{abstract}
An analytically tractable model for Gaussian multiuser channels with fading is studied, and the capacity region of this model is found to be a good approximation of the capacity region of the original Gaussian network. This work extends the existing body of work on deterministic models for Gaussian multiuser channels to include the physical phenomenon of fading. In particular, it generalizes these results to a unicast, multiple node network setting with fading.
\end{abstract}

\section{Introduction}
As capacity results for Gaussian multiuser networks are in general difficult to obtain, meaningful models that capture the capacity trends of these networks are very useful. Recently, seminal work in this domain by Avestimehr, Diggavi and Tse~\cite{tse1,tse2,tse3} has resulted in deterministic models which are easier to analyze that the original Gaussian network and can be shown through examples to approximate the actual capacity of the channel fairly well.  A bound of the difference between the capacity of the deterministic model and the general, Gaussian unicast network has also been found~\cite{tse4}.  The core idea is the representation of the channel in terms of a deterministic input and output alphabet relationship that reflects the signal to noise ratio (SNR) at each node in the network.

The goal of this paper is to introduce fading into this modeling framework. In general, fading, modeled in its simplest form as a multiplicative channel state, adds an additional dimension of complexity to a capacity problem. There are Gaussian channels where capacity without fading is known but with fading unknown (an example is the fading broadcast channel where the transmitter does not know the state). Thus, analytically tractable models that can, with a fair degree of accuracy, capture fading in Gaussian channels can prove very useful in capacity characterizations for Gaussian networks with fading. This paper assumes that, in each case, only the receiver(s) know the fading state and the transmitter(s) do not.

We introduce the term ``quasi-deterministic network'' in this paper, to describe most generally, any network which is deterministic, given some random state variable $S$ which is independent of all inputs. In this paper, $S$ is $iid$ over each timestep. The network models studied in each of the papers \cite{tse4},\cite{DGPHE:2006}, and \cite{brian} are all examples of quasi-deterministic networks.

This paper has a relatively straightforward progression. The next section describes the quasi-deterministic model presented in this paper using the point-to-point channel, and summarizes the main results obtained for different multiuser channels. In Section \ref{sec:mac}, a closed-form expression for the capacity region of the multiple access channel (MAC) is derived, and the model is compared to the Gaussian case. Section \ref{sec:broadcast} illustrates the case of the semi-deterministic broadcast channel. Section \ref{sec:unicast} demonstrates that the cut-set bound on capacity of a unicast network of such fading channels, and in fact any quasi-deterministic network, is achievable when the fading state $S$ is available to the final destination.  

\section{Model for Fading Gaussian Channels}
\label{sec:model}

\subsection{Notation}

For a vector $X$ of length $n$ denote by $X^i$ the $i^{th}$ most significant bit, i.e. $X^1$ is the most significant bit and $X^n$ is the least significant. Also, $\lg$ denotes logarithm base $2$. For addition, ``$\oplus$'' is the bit-level by bit-level finite-field summation of two vectors, whereas  ``$+$'' is the algebraic addition of two signals. For a matrix, ``rank'' is the rank, i.e. the number of linearly independent rows (or columns).

\subsection{Model}

The simplified model that we introduce for fading Gaussian channels is based on the work on deterministic modeling of Gaussian channels introduced in~\cite{tse1}, and is similar to the model presented in \cite{tyz08}.  For motivation, and to capture the spirit of the modeling assumptions, we briefly describe the translation of the point-to-point fading Gaussian channel to our quasi-static model.

In~\cite{tse1}, the case of a real AWGN channel with unit noise power and unit power constraint, i.e. $Y = HX + Z$ where $E[X^2]\le1$ is the average power constraint and $Z\sim\mathcal{N}(0,1)$, is considered.  The capacity of such a channel, $\frac{1}{2} \lg(1+SNR)$ can be approximated as $\lg\sqrt{SNR}=\lg H$.  Thus, the paper intuitively models a point-to-point Gaussian channel as a pipe which truncates the transmitted signal and only passes the $\lg H$ bits which are above noise level. The point-to-point Gaussian channel has thus been modeled as a bit pipe which transmits some number $m$ of the most significant bits of the input, where $m=\lceil{\frac{1}{2}\log SNR}\rceil$ for real signals.

This paper takes a similar approach to modeling fading channels.  As in~\cite{tse1}, the input to our point-to-point channel model, $X$, will consist of a vector of fixed length $n$ bits.  The output of the channel $Y$ at time $t$ will consist of a vector of length $m(t)$ bits.  The effect of receiver fading is modeled as the random variation in $m(t)$ over time, which is denoted by the random variable $M$.  The number of (most significant) bits received (which is a realization of $M$) is determined by the fading and is independent of the input $X$ and known only at the receiver.
The number of received bits $M$ is a random variable which takes on integer values $0\le m(t) \le n$:  say $P\left[m=i\right]=p_i$.
\begin{figure}[ht]
\center
\includegraphics[scale=0.25]{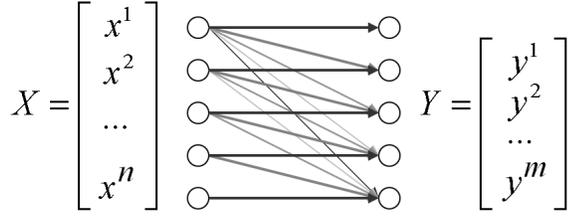}
\caption{Model for the point-to-point channel} \label{fig:p2p1}
\end{figure}

That is, if $x(t)=\left \{x^i(t):i\in\{1,..,n\}\right\}$, then $y(t)=\{x^i(t)|i\in\{1,...,m(t)\}\}$ where $m(t)$ is the realization of the fading random variable $M$.

The capacity of this model for the fading point-to-point channel is therefore
\begin{align*}
I(X;Y,M) = I(X;M) + I(X;Y|M).
\end{align*}
Since $X$ and $M$ are independent, the first term is zero; a uniform binary input for $X$ maximizes the second term as $E[M]$, that is, the average number of bits seen by the receiving node. In fact,
\begin{eqnarray}
\label{eq:capacity}
I(X;Y|M) & = & H(Y|M)-H(Y|X,M) \\
& = & H(Y|M) \nonumber\\
& = & \sum Pr(M=m) * H(Y|M=m) \nonumber\\
& = & E\left[M\right]
\end{eqnarray}
where (\ref{eq:capacity}) comes from the fact that $Y$ is a deterministic function of $X$ and $M$.  Intuitively, this result corresponds to that of the fading Gaussian point-to-point channel, with capacity $E[\frac{1}{2}\lg(1+SNR)]$. Figure~\ref{fig:p2p1} illustrates the point-to-point model. An $n$-bit vector is truncated into an $m$-bit vector depending on the realization $m$ of the fading random variable $M$. The main difference between the model and the fading Gaussian is that $M$ has integer realizations and $\lg(1+SNR)$ has in general, real valued realizations. Thus, some difference or "loss" corresponds to the integer truncation of each rate term. Therefore, for high $SNR$ ($SNR \ge 1$), we can write:
\begin{align}
\label{eq:diff}
\left| E\left[M\right] - \left[\frac{1}{2}\log (1+SNR)\right] \right| \leq 1
\end{align}

\section{Multiple Access Channel}
\label{sec:mac}

\begin{figure}[ht]
\centering
\subfigure[Model for MAC]{
\includegraphics[scale=0.3]{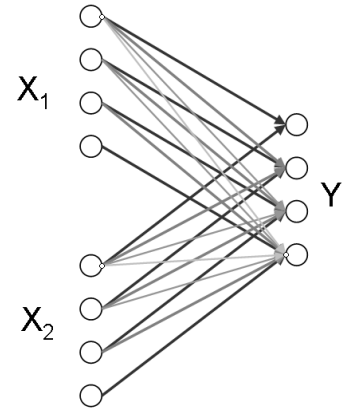}
\label{subfig:mac1}
}
\subfigure[Difference between model and Gaussian]{
\includegraphics[scale=0.35]{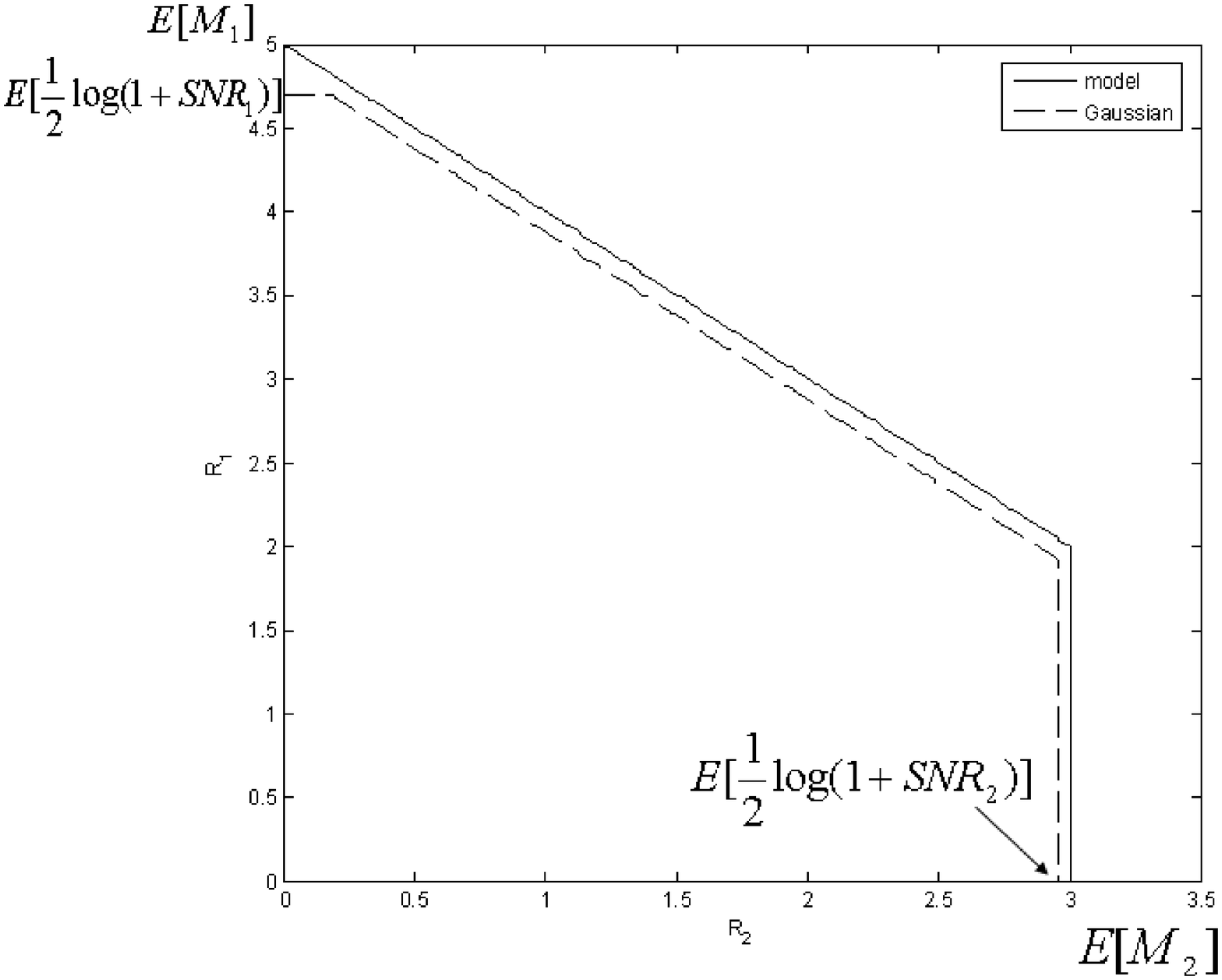}
\label{subfig:mac2}
}
\caption{\subref{subfig:mac1} Model for MAC  \subref{subfig:mac2} Comparison with the Gaussian MAC capacity} \label{fig:mac11}
\end{figure}

In a the two-user Gaussian fading MAC channel, the received signal is given by
\begin{align*}
Y=H_1X_1 + H_2X_2 + Z,
\end{align*}
where $Z\sim\mathcal{N}(0,1)$, $H_1 \ge 0$ and $H_2 \ge 0$ are the fading channel gains. We assume $SNR_2 \le SNR_1$ without loss of generality. For the model depicted in Figure~\ref{subfig:mac1}, we define the number of bit-levels randomly received from user $k$ at the receiver by $M_k$.  The receiver knows both fading states $M_k$. The two inputs to this MAC are the $n_{k}$ length vectors $X_k(t)=\left \{x^i_k(t):i\in\{1,..,n_{k}\}\right\}$, $k\in\{1,2\}$, while the single output $Y$ is a vector bit-level by bit-level finite-field summation of $X_1$ and $X_2$, appropriately shifted by the fading levels $M_k$. Specifically, denoting by $y^i(t)$ the $i^{th}$ most significant bit in the bit-level expansion of the vector $y(t)$, we can write $y^i(t)$ as

\begin{eqnarray}
y^i(t)=\left\{x^{i-\eta_1(t)}_1+x^{i-\eta_2(t)}_2(t):i\in\{0,\dots,\hat{m}(t)\}\right\}
\end{eqnarray}
where $\hat{m}(t)=\max(m_1(t),m_2(t))$, $\eta_1(t) = (\hat{m}-m_1)(t)$, $\eta_2(t) = (\hat{m}-m_2)(t)$ and we set $x_k^j=0$ for $j\le 0$.

The capacity region of the MAC channel is therefore given by
\begin{eqnarray}
R_1 &\le& E\left[M_1\right]\\
R_2 &\le& E\left[M_2\right]\\
R_1 + R_2 &\le& E\left[\max(M_1,M_2)\right]
\end{eqnarray}
Figure~\ref{subfig:mac2} illustrates the capacity region of this model and compares it to a simulated Gaussian case where $SNR_{\max} = SNR_1 > SNR_2$, $E\left[M_1\right] = E\left[\lceil{\frac{1}{2}\log(1+SNR_1)}\rceil\right]=5$ and $E\left[M_2\right]=3$.

The achieved capacity is at most within 1.5 bits from that of the Gaussian MAC with fading. In fact,
\begin{eqnarray}
\label{eq:maceq}
R_1 &\le& E[\frac{1}{2}\log(1+SNR_1)] \nonumber\\
R_2 &\le& E[\frac{1}{2}\log(1+SNR_2)] \nonumber\\
R_1 + R_2 &\le& E[\frac{1}{2}\log(1+SNR_1+SNR_2))] \nonumber\\
&\le& E[\frac{1}{2}\log(1+2SNR_{max}))]\nonumber\\
& \le & E[\frac{1}{2}\log(1+SNR_{max}))]+\frac{1}{2}.
\end{eqnarray}

The model hence gives a good approximation of the Gaussian MAC channel under the presented fading model. It can be seen from Equation~\ref{eq:maceq} that the capacity of this model lies within 1.5 bits of the Gaussian MAC capacity.


\section{Broadcast Channel and the Capacity of the Semi-Deterministic Broadcast Case}
\label{sec:broadcast}

Since the capacity of the fading Gaussian broadcast channel is yet unknown, a model for the corresponding simplified channel model can serve two purposes. First, it may help us benchmark the performance of practical wireless communication systems with fading. Second, it may suggest achievable schemes for the original Gaussian fading broadcast channel.

The input $X$ for the fading broadcast channel model will consist of a vector of a fixed number $n$ bits.  Receiver $1$ sees the $m_1(t)$ most significant bits of the input, while Receiver $2$ sees the $m_2(t)$ most significant bits. The values $m_1(t)$ and $m_2(t)$ are realizations of the independent random variables $M_1$ and $M_2$ and are known to the their respective receivers, only.

In~\cite{tyz08}, Yates et al. find an achievable region for the fading broadcast channel, that lies within a constant gap of $1.44$ bits/s/Hz of the capacity region.

We now turn our attention to the semi-deterministic case, where we determine capacity in the hope of finding better achievable schemes to approximate the capacity of the one-sided fading Gaussian broadcast channel.
Note that a single letter characterization for semi-deterministic channels is known, but here we use the K\"{o}rner-Marton outer bound as our starting point for the analysis (which is tight on the capacity region of semi-deterministic channels).  The motivation for this is to shed light on the choice of auxiliary random variable $V$ which motivates one particular coding scheme that achieves capacity.

The semi-deterministic broadcast model studied here can be summarized by the expressions
\begin{align}
\label{eq:semibc}
Y_1 = \left[X^1 \ldots X^{m_1}\right] & & Y_2 = \left[X^1 \ldots X^{M_2}\right]
\end{align}
with input $X=\left[X^1 \ldots X^{n}\right]$, where $m_1$ constant with

$0<m_1<n$, $M_2 \sim p(i)$ with $\mathcal{M}_2 = \{0,1,\ldots,n\}$.

For the channel model described in Equation (\ref{eq:semibc}), we first show that the K\"{o}rner-Marton outer bound ~\cite{Marton} (equivalently, semi-deterministic capacity region) for this broadcast channel is easy to evaluate, and then show that it is achievable using superposition coding.  Note that, for a general semi-deterministic channel, superposition coding is not sufficient to achieve capacity.

\subsection{Converse}
Note that the boundary defined by the following optimization problem
\begin{equation}
\label{eqn:opt}
\max_{p(v,x)} I(X;Y_1|V) + \mu I(V;Y_2)
\end{equation}
for all $\mu \ge 0$  is an outer bound on the  K\"{o}rner-Marton region~\cite{Marton}, and thus we focus on  this optimization problem instead.

Because the receiver has access to the channel state,
\begin{eqnarray}
I(V;Y_2) & = & I(V;Y_2,M_2) \nonumber\\
& = & I(V;M_2) + I(V;Y_2|M_2) \nonumber\\
& = & I(V;Y_2|{M_2}) \nonumber
\end{eqnarray} as $V$ and $M_2$ are independent. Thus, the optimization problem in \ref{eqn:opt} translates into
\begin{eqnarray}
\label{eq:finalopt}
\max_{p(v,x)} & H(Y_1|V) - \mu H(Y_2|V,M_2) + \mu H(Y_2|M_2) \nonumber\\
\max_{p(v,x_1,\ldots,x_{n})} & H(X^1 \ldots X^{m_1}|V) - \mu \times \nonumber\\
& \sum_{i=0}^{n} p(i) \left[H(X^0 \ldots X^i|V) - H(X^0 \ldots X^i)\right] \nonumber\\
\max_{p(v,x_1, \ldots, x_{m_1})} & \sum_{j=1}^{m_1} H(X^j|V,X^0,\ldots,X^{j-1}) (1 -\mu q(j)) \nonumber\\
& +\mu q(j) H(X^j| X^0, \ldots, X^{j-1}) \nonumber\\
& +\mu \sum_{j=m_1+1}^{n} p(j) I(V;X^j|X_0,\ldots,X^{j-1})
\end{eqnarray}
where $X^0 = \phi$, $q(j) = \sum_{i=j}^{m_1} p(i)$ and is thus a non-decreasing sequence.

\begin{figure}[ht]
\begin{center}
\scalebox{0.7}{\input{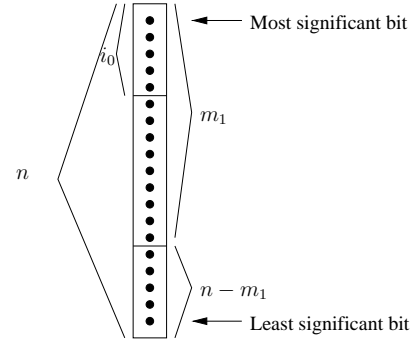}}
\caption{Relationship of $U$ and $V$ to $X$} \label{fig:bits}
\end{center}
\end{figure}

Let $i_0$ be such that $q(i_0+1) < 1/\mu$ and $q(i_0-1) \ge 1/\mu$. It is clear that choosing $X^i$s independent maximizes the objective in (\ref{eq:finalopt}). In addition, $V$ must include the following two components:
$\left[X^{0} \ldots X^{i_0}\right]$ the first $i_0$ components of the input and $\left[X^{m_1+1} \ldots X^{n}\right]$ the last $(n-m_1)$ components of the input (that are never received by Receiver 1).  This assignment is illustrated in Figure \ref{fig:bits}.

\subsection{Achievability and Discussion}
\label{sec:ach}
The converse helps determine what form the auxiliary random variables $U$ and $V$ should take in the achievability argument.  We have Marton's Inner Bound~\cite{Marton}:

\begin{align*}
R_1&\le I(U;Y_1)\\
R_2&\le I(V;Y_2)\\
R_1+R_2&\le I(U;Y_1)+I(V;Y_2)-I(U;V)
\end{align*}
for some $p(u,v,x)=p(u,v)p(x|u,v)$.

Choose any integer $i_0$ such that $0\le i_0 \le m_1$, and let $U$ and $V$ be uniformly binary random vectors of length $(m_1-i_0)$ and $i_0 + (n-m_1)$, respectively.  The $n$ length (uniformly distributed) binary vector $X$ is formed by concatenating first $i_0$ bits of $V$ (in the most significant positions of $\tilde{X}$), then the $(m_1-i_0)$ bits of $U$, and finally the remaining $(n - m_1)$ bits of $V$ in the least significant positions.

From this choice of auxiliary random variables, it is clear that $I(U;V)$ is zero, $I(U;Y_1)$ is $m_1-i_0$ and
\begin{align*}
I(V;Y_2) = \sum_{i=1}^{i_0} ip(i) + \sum_{i=m_1+1}^n \left(i_0 + (i-n)\right)p(i).
\end{align*}
Intuitively, this strategy has a straightforward implication.  Since the lowest level bits are never received by Receiver $1$, they should always be assigned to Receiver $2$.  If the user desires to dedicate more bits to Receiver $2$, it is immaterial to Receiver $1$ which bits he chooses, since each contributes an equal amount of rate.  However, to maximize the amount of data that can be transmitted to the second receiver, the user should first assign the bits which are most likely to be received (specifically, the most significant bits) to Receiver $2$ before any others.  Also note that this achievability can also be easily generalized to the two-sided fading broadcast channel. In fact, this coding scheme and observations were also made by Yates and Tse in~\cite{tyz08}.

\section{General Unicast Network}
\label{sec:unicast}

We consider a general unicast network $\G$ of fading channels with each channel modeled as in Section \ref{sec:model} and having broadcast and multiple access properties. The network is a directed graph $\G = (\mathcal{V}, \mathcal{E})$, where each node $j \in \mathcal{V}$ has some power and therefore can transmit the symbol $X_j(t)=\{x_j^k(t) | k=1,\dots,l\}$, i.e. each symbol has $l$ bit levels. Note that $k=1$ is the most significant bit. In this scheme, symbol fading or fast fading is assumed, and all the fading states are known to the ultimate destination and to the respective receivers in each transmission. This network is actually a particular case of a quasi-deterministic network which we define next.


\subsection{Quasi-Deterministic Networks}

A quasi-deterministic network is a general network in which the channel model with input $x$, output $y$ and state $s$ is given by $p(x,y,s)=p(s) \times p(x) \times p(y|x,s)$ where $p(y|x,s)$ is a deterministic function and $x$ is independent of $s$. Fading state $S$ is a random variable which is $iid$ for each timeslot in this work, i.e. fast fading is assumed.

\subsection{Network Model}
\label{subsec:netmodel}

The network model studied in this paper is the linear finite-field deterministic model presented in~\cite{tse1}, augmented with fading as explained in Section~\ref{sec:model}. This network is a particular case of the quasi-deterministic network. Here, $\G$ is a directed acyclic graph.  Then, every node $j$ has a number $l$ of bit-levels, and each bit-level receives the finite-field sum in $GF(2)$. In other terms, the signal received at a node j, similarly to the signal in Section~\ref{sec:mac}, is given by

\begin{equation*}
y_j^k(t)=\sum_{r \in \mathcal{N}_I(j)} \{x^{k-(\hat{m}-m_r)(t)}_r(t) | k \in {0,\dots,\hat{m}(t)}\}
\end{equation*}
where $\mathcal{N}_I(j)$ is the set of nodes with edges incident on node $j$, $\hat{m}(t) = \max_{r \in \mathcal{N}_I(j)} m_r(t)$ and $m_r(t)$ is the fading realization for edge $(r,j)$ at time $t$, and the summation is of the type $\oplus$.

It is useful to note a difference between the model presented here and the model given by Avestimehr et al. in ~\cite{adt08}, where the channel gains are also chosen from a set for each link, however the fading state distribution is unknown at the sender. In this paper, we assume that the distribution of the fading state is known at the sender and therefore we can achieve a rate better than the $inf$ of the cut-set bound in ~\cite{adt08}, i.e. the worst case. In fact, it turns out that the average value of the cut-set bound is achievable.

\subsection{Upper Bound}
\label{subsec:upperbound}

Let $\mathcal{V}$ be the set of vertices of $\G$, $S$ a random vector of size $|\xi|$, where $|\xi|$ is the number of edges of $\G$. $S$ is a collection of all the state random variables in the network for a particular timeslot. For a quasi-deterministic network, $S$ can be thought of as the state of the network at each time instant. The set of all cuts of the network is denoted by $\Lambda_D$. For the special case of this fading network, we define, similarly to $G_{\Omega,\Omega^C}$ \cite{tse1}, $A_{S,\Omega}$ to be the random total transfer matrix associated with a cut $\Omega \in \Lambda_D$, i.e. the relationship between the concatenated signal $X_c$ sent by the nodes on the left side of the cut and the resulting signal $Y_c$ received by the nodes on the right side of the cut is $Y_c=A_{S,\Omega}X_c$.

The randomness of the matrix $A_{S,\Omega}$ is a result of the randomness of the random vector $S$, for a fixed cut $\Omega$.
Now, using the general cut-set upper bound for a general network, we can write by~\cite{ElementsInfoTheory} and~\cite{tse2},
\begin{equation}
\label{eq:upperbound}
R \le \displaystyle\max_{p(x_1,\dots,x_{|\mathcal{V}|})} \displaystyle\min_{\Omega \in \Lambda_D} I(X_\Omega;Y_{\Omega^C}, S|X_{\Omega^C})
\end{equation}
In fact, for the particular fading model studied in this paper (model in \ref{subsec:netmodel}),

\begin{eqnarray}
I(X_\Omega;Y_{\Omega^C}, S|X_{\Omega^C}) & = & I(X_\Omega;Y_{\Omega^C}|X_{\Omega^C}, S) \nonumber\\
& = &H(Y_{\Omega^C}|X_{\Omega^C}, S) \nonumber\\
\label{eq:quasicut}
& = &\mathbb{E}_S H(Y_{\Omega^C}|X_{\Omega^C}, S=s) \\
\label{eq:fadingcut}
& = &\mathbb{E}_S \text{rank}(A_{S,\Omega})
\end{eqnarray}

where (\ref{eq:quasicut}) is the cut-set upper bound for the general quasi-deterministic network and (\ref{eq:fadingcut}) is its particular value for our fading network model, where $A_{S,\Omega}$ is the transfer matrix for a certain cut $\Omega$.

\section{Achievability in Quasi-deterministic Unicast Networks with Random Coding}
\label{sec:random}

The goal now is to show that, using random coding, we can achieve rates arbitrarily close to the upper bound specified in \ref{subsec:upperbound} for the network model in~\ref{subsec:netmodel}. Also, the bound given by Theorem \ref{thm:bound} is achievable for quasi-deterministic networks.

\begin{thm}
\label{thm:bound}
Given a quasi-deterministic unicast network with the model specified in Section~\ref{sec:unicast}, the rate given by
\begin{equation}
R \le \displaystyle\max_{\prod p(x_j), j \in \mathcal{V}} \displaystyle\min_{\Omega \in \Lambda_D} \mathbb{E}_S H(Y_{\Omega^C}|X_{\Omega^C}, S)
\end{equation}
is achievable, and is equivalent to the upper bound given by \ref{eq:upperbound} for the fading network, i.e. for the fading network model defined in \ref{subsec:netmodel}, \ref{eq:upperbound} and \ref{eq:fadingcut} are equivalent. Here, $\Omega$ is a cut, and $\Lambda_D$ the set of all cuts.
\end{thm}

To prove this, we need to prove that the upper bound in Section~\ref{subsec:upperbound} is achievable. We will proceed along similar lines to the proofs in~\cite{brian} and~\cite{tse2} and use random coding arguments to get the result.

Let $\mathcal{W}=\{1,2,\dots,2^{nRB}\}$, where $R$ is the desired rate, $B$ the number of blocks to send and $n$ the block size. If $L$ is the longest path in the network, the transmission will take place in $(B+L)n$ timeslots, achieving a rate of $R \times \frac{B}{B+L}$, which approaches $R$ as $B$ gets large.

\subsection{Encoding and Decoding}

As in~\cite{brian}, each node $i$ generates $(B+L)$ codebooks, where each codeword is $nl$ bits long, $l$ being the number of levels at each node and codewords are all generated with the distribution $\prod_1^{n} p(x)$ where X is a Bernoulli($1/2$) random vector of size $l$. The final destination knows all codebooks and all the states $S^{n(B+L)}$ of the network during transmission time.
Denoting by $x_i(b)$ the transmitted signal of node $i$ during the transmission of block $b$, $x_i(b) = f_i^{(b)} (y_i(b-1))$ where $y_i(b-1)$ is the block received on $i$'s incoming edge during transmission time of block $(b-1)$, and $f_i^{(b)}$ is the random function chosen at each block period for every outgoing edge of node $i$.

To decode the message, the destination node deterministically simulates all the $2^{nRB}$ messages, knowing all the fading states and all the codebooks used during transmission time. If the output observed when simulating exactly one $w \in \mathcal{W}$ is identical to the actual signal, then $w$ was transmitted, otherwise an error is declared. Thus, an error occurs if the fading pattern is not typical or if two codewords produce the same output at the destination node, which we shall detail next.

\subsection{Probability of Error Calculation}
\label{subsec:proberror}

An error occurs at the destination node if the fading is not typical, the probability of which can be made small when a large enough $n$ is chosen. Let us turn our attention to the error event where two codewords produce the same output, which is more involved. Suppose that codeword $w_1$ is transmitted. Define $E_j$ to be the event that codewords $w$ and $w_j$ produce the same output after the simulation of the network by the destination node. Then the error event associated with transmitting $w$ is

\begin{center}{$E = \bigcup_{j=1}^{2^{nRB - 1}} E_j$}\end{center}
Let $\mathcal{V}_s$ and $\mathcal{V}_d$ denote the nodes on the source and the destination side, respectively. As in~ \cite{brian}, define, for a cut $\Omega$, $F_\Omega^b$ as the event that after the $b^{th}$ block is simulated, the inputs to all the nodes in $\mathcal{V}_d$ are identical and at least two of the inputs of the nodes in $\mathcal{V}_s$ are different. So, if $w$ and $w_1$ produce the same inputs at the destination node, one of the events $F_\Omega^b$ has occurred. So we can write $E_1$ as

\begin{eqnarray}
E_1 = \bigcup_{(\Omega_{1}, \Omega_{2}, \dots, \Omega_{B+L}) \in \Omega_D} \big{(F_{\Omega_1}^1 \cap F_{\Omega_2}^2 \cap \dots \cap F_{\Omega_{B+L}}^{B+L}\big)} \nonumber
\end{eqnarray}
where $(\Omega_1, \Omega_2, \dots, \Omega_{B+L})$ is a sequence of cuts corresponding to transmission times of blocks $1,2,\dots,(B+L)$ and $\Omega_D$ is the set of all sequences of cuts. To calculate $Pr(E_1)$, we will use the union bound for all sequences of cuts.

Note in this case that the event $F_{\Omega_k}^k$ is only dependent on the event $F_{\Omega_{k-1}}^{k-1}$, since random coding is performed independently on each outgoing edge and for each block. We assume the final destination knows all the fading realizations in the $n(B+L)$ timeslots. Using the worst-case cut sequence and the union bound over all possible sequences of cuts, and denoting by $k$ the total number of sequences of cuts (which is finite), we can then write


{\setlength\arraycolsep{2pt}
\begin{eqnarray}
\lg Pr(E_1) &\le& \lg k + \lg Pr(F_{\Omega_1}^1) + \lg Pr (F_{\Omega_2}^2|F_{\Omega_1}^1) \nonumber\\
& & + \dots + lg Pr(F_{\Omega_{B+L}}^{B+L}|F_{\Omega_{B+L-1}}^{B+L-1}) \nonumber \\
& \le & lg k + \lg (1) + \lg Pr (F_{\Omega_2}^2|F_{\Omega_1}^1) \nonumber\\
& & + \dots + \lg Pr(F_{\Omega_{B+L}}^{B+L}|F_{\Omega_{B+L-1}}^{B+L-1}) \nonumber \\
& = & \lg k - n H(Y_{\Omega_2^C}|X_{\Omega_1^C}, S) \nonumber\\
& & - \dots -n H(Y_{\Omega_{B+L}^{C}}|X_{\Omega_{B+L}^C}, S)\\
& = & \lg k - n \displaystyle\sum _{i=2}^{B+L} H(Y_{\Omega_i^C}|X_{\Omega_{i-1}^C}, S)
\end{eqnarray}}

Now using lemma $6.4$ and the proof of lemma $6.2$ from~\cite{tse2}, we have that for any $i$,
\begin{eqnarray}
\displaystyle\sum_{i=2}^{B+L} H(Y_{\Omega_i^C}|X_{\Omega_{i-1}^C}, S) &\ge& (B+L-2^{|\mathcal{V}|-2}+1) \times \nonumber\\
& & \displaystyle\min_{\Omega \in \Lambda_D} H(Y_{\Omega^C}|X_{\Omega^C}, S) \nonumber
\end{eqnarray}

$\lg Pr(E_1)$ can now be upper bounded by
\begin{eqnarray}
\lg Pr(E_1) &\le& \lg k -n (B+L-2^{|\mathcal{V}|-2}+1) \times \nonumber\\
& & \displaystyle\min_{\Omega \in \Lambda_D} H(Y_{\Omega^C}|X_{\Omega^C}, S) \nonumber
\end{eqnarray}
Using the union bound for the probability of error we get
\begin{eqnarray}
\lg Pr(E) &\le & n R B + \lg k - n (B+L-2^{|\mathcal{V}|-2}+1) \times \nonumber\\
& & \displaystyle\min_{\Omega \in \Lambda_D} H(Y_{\Omega^C}|X_{\Omega^C}, S) \nonumber \\
&\le& \lg k + n B (R - \displaystyle\min_{\Omega \in \Lambda_D} H(Y_{\Omega^C}|X_{\Omega^C}, S) - \epsilon ) \nonumber
\end{eqnarray}
where the last inequality is obtained for $B$ large enough. Hence, for $R \le \displaystyle\min_{\Omega \in \Lambda_D} H(Y_{\Omega^C}|X_{\Omega^C}, S)$, $\lg Pr(E) \to -\infty$ $Pr(E) \to 0$ and the rate in~\ref{eq:upperbound} is achievable.
In the particular case of the fading network, $H(Y_{\Omega^C}|X_{\Omega^C}, S)$ evaluates to $\text{rank}(A_{S,\Omega})$, as mentioned in~\cite{tse1} and hence the result in (\ref{eq:fadingcut}).

\section{Conclusion}
\label{sec:conc}

In this paper, an equivalent quasi-deterministic model of the Gaussian channel was presented, along with the comparison to the original Gaussian channel in the fading point-to-point, MAC and semi-deterministic broadcast case. For the general unicast network, it was proven that the min cut is achievable for the quasi-deterministic network model using random coding. Combining our result with the result of~\cite{tse4} shows that we can find the capacity of the corresponding Gaussian network to within a constant bound independent of the channel parameters, similarly to~\cite{adt08}.

\bibliographystyle{IEEEtranS}
\bibliography{feedback}
%
%
\end{document}